\documentclass[runningheads]{llncs}
\usepackage{ctable}
\usepackage{xcolor}
\usepackage{graphicx}
\usepackage{mathtools}
\usepackage{multirow}
\usepackage{algorithm}
\usepackage{algorithmic}
\usepackage{adjustbox}
\begin{document}
\title{Energy Efficient LSTM Accelerators for Embedded FPGAs through Parameterised Architecture Design}
\titlerunning{Energy Efficient LTSM Accelerators for Embedded FPGAs}
%
%

\author{Chao Qian\inst{}\orcidID{0000-0003-1706-2048} \and
Tianheng Ling\inst{}\orcidID{0000-0003-4603-8576} \and
Gregor Schiele\inst{}\orcidID{0000-0003-4266-4828}}
\authorrunning{C. Qian et al.}
\institute{Embedded Systems Lab, University of Duisburg-Essen \\ Duisburg, Germany\\
\email{\{chao.qian, tianheng.ling, gregor.schiele\}@uni-due.de}}



%
\maketitle              
\begin{abstract}

Long Short-term Memory Networks (LSTMs) are a vital \textit{Deep Learning} technique suitable for performing on-device time series analysis on local sensor data streams of embedded devices. In this paper, we propose a new hardware accelerator design for LSTMs specially optimised for resource-scarce embedded Field Programmable Gate Arrays (FPGAs). Our design improves the execution speed and reduces energy consumption compared to related work. Moreover, it can be adapted to different situations using a number of optimisation parameters, such as the usage of DSPs or the implementation of activation functions. We present our key design decisions and evaluate the performance. Our accelerator achieves an energy efficiency of 11.89 GOP/s/W during a real-time inference with 32873 samples/s.

\end{abstract}

\section{Introduction}

Recent studies have shown the superiority of \textit{Deep Learning} algorithms over traditional methods for time series analysis \cite{lim2021time,lara2021experimental}. Among these algorithms, Long Short-term Memory Networks (LSTMs) have been extensively studied for their ability to model and predict nonlinear time-varying systems \cite{lindemann2021survey}. Running LSTMs at the edge, especially on embedded sensor devices, is preferable for tasks with data privacy and security requirements, such as data collection at public locations \cite{huang2018deep}. In addition, on-device inference with low latency is critical for many applications, like human voice analysis with wearable devices \cite{conti2018chipmunk}. However, deploying LSTMs on devices faces challenges due to limited local computational resources and energy. Microcontrollers are often not fast enough, while GPUs consume too much energy.
One promising approach is to design LSTM accelerators for Field-Programmable Gate Arrays (FPGAs), which offer fast computation and reconfigurability while being typically more energy-efficient \cite{chen2019deep}. This paper proposes a novel LSTM accelerator architecture for embedded FPGAs. We use an Xilinx \textit{Spartan-7} \textit{XC7S15} FPGA. Our main contributions are as follows:

\begin{itemize}
    \item Our LSTM accelerator architecture achieves superior resource utilisation compared to state-of-the-art approaches. We accomplish this by using more efficient activation functions and quantising to 8 bits. We achieve an average reduction of 29.62\% in LUT utilisation and 33.33\% in LUTRAM utilisation.
    
    \item Our design offers the option of not using DSPs for arithmetic logic to overcome the limitations of prior work that heavily relies on DSPs. This way, we can support LSTM models with up to 5 LSTM layers, each of which can have a maximum hidden size of 60. 

    \item We significantly reduce the logic and net delay in the LSTM accelerator by optimising the activation function and Arithmetic-Logic Unit (ALU) implementation. The accelerator's maximum clock frequency increases to 204 MHz, leading to nearly a 2$\times$ increase in throughput.
    
    \item We validate our proposed architecture by implementing it in Vivado and real hardware. Our results demonstrate a reduced power consumption of up to 18.57\% and an improved energy efficiency per inference of 59.19\%.
\end{itemize}

In the remainder of this paper, we first discuss related research in Section 2. Then, Section 3 provides background information on LSTMs. Our design is described in Section 4, while Section 5 presents implementation details. An evaluation of our work is conducted in Section 6. Section 7 concludes the paper and outlines future research plans.
\section{Related Work}

Numerous studies have investigated the design of LSTM accelerators for FPGAs, but most research has concentrated on either server-grade FPGAs installed in the Cloud \cite{boutros2020beyond} or mid-range FPGAs in Edge servers \cite{cao2019efficient,varadharajan2022p,zhang2017power}. To our knowledge, only a few papers have discussed the design of LSTM accelerators for embedded FPGAs. Due to their low cost, compact size, and low power consumption, such FPGAs can provide flexible hardware acceleration for embedded devices, e.g., in the Internet of Things. However, they have far fewer resources (in terms of LUTs, DSPs, RAM, etc.) than bigger FPGAs, requiring compact accelerator designs. In addition, such accelerators must be optimised for energy efficiency to not limit the lifetime of battery-operated devices. To achieve the required performance while adhering to size and power limits, careful study and hardware resource optimisation are necessary to design LSTM accelerators on embedded FPGAs.

According to a study by Hasib-Al-Rashid et al. \cite{manjunath2020low}, the static power consumption of the \textit{Artix 7} \textit{XC7A100T} FPGA has a significant negative impact on the overall energy efficiency of their LSTM accelerators. One possible solution to mitigate this issue is to use FPGAs with negligible static power consumption. For instance, Chen et al. \cite{chen2021eciton} implemented their LSTM accelerator on the \textit{iCE40} UltraPlus \textit{UP5K} FPGA, which has a static power consumption at the $\mu$A level, resulting in an energy efficiency reported to be $7.6\times$ better than that of \cite{manjunath2020low}. However, the low maximum clock frequency (17MHz) of the chosen FPGA limited the maximum throughput of the accelerator to 0.067 GOP/s, which could pose challenges in supporting real-time inference applications. Furthermore, their accelerator implemented an LSTM model with a single LSTM cell but already occupies 75\% of DSPs, 100\% of SPRAM, 73.3\% EBR-RAM and 94.5\% of LUTs. This makes scaling up to bigger LSTM models impossible.

In 2022, Qian et al. \cite{qian2023enhancing} proposed an approach to reduce the proportion of static power in the overall power consumption of the \textit{Spartan-7} \textit{XC7S15} FPGA. They achieved this by increasing the dynamic power consumption through parallelism in the LSTM cell, resulting in a throughput of 0.363 GOP/s at 100 MHz. This throughput is 5.4$\times$ faster than the maximum throughput achieved by the approach proposed by Chen et al. \cite{chen2021eciton}. In addition to the higher throughput, Qian et al. achieved 1.37$\times$ better energy efficiency with 5.33 GOP/J compared to Chen et al.'s approach.

While previous studies have shown promising results in accelerating LSTM models, there are still limitations concerning the scalability of the FPGA and its maximum usable clock frequency. For example, the FPGAs from the \textit{Spartan-7} family can implement fixed-point arithmetic at frequencies up to 239MHz \cite{burger2020architecture}, which presents an opportunity for optimising the accelerator for higher operating frequencies and better energy efficiency. Therefore, further research is necessary to identify and develop optimisations for more scalable and energy-efficient LSTM accelerators for embedded FPGAs.

\section{LSTM Background}\label{sec:background}

This section presents the fundamental concepts of LSTMs necessary to understand our proposed architecture design. For simplicity, we use a basic LSTM model specifically developed for single-step ahead time series prediction. The model comprises an LSTM layer with one LSTM cell, followed by a dense layer, as detailed in \cite{fu2016using}. We assume that the input sequence $X=\{x_{t-N+1}, \ldots, x_{t-1}, x_{t}\}$ has length $N$ and each element is of $M$ dimensions, where $M\geq 1$ to support both univariate and multivariate time series. 
The input sequence $X$ is iteratively processed through an LSTM cell within the LSTM layer.

\begin{figure}[htb]
    \centering
    \includegraphics[width=.9\textwidth]{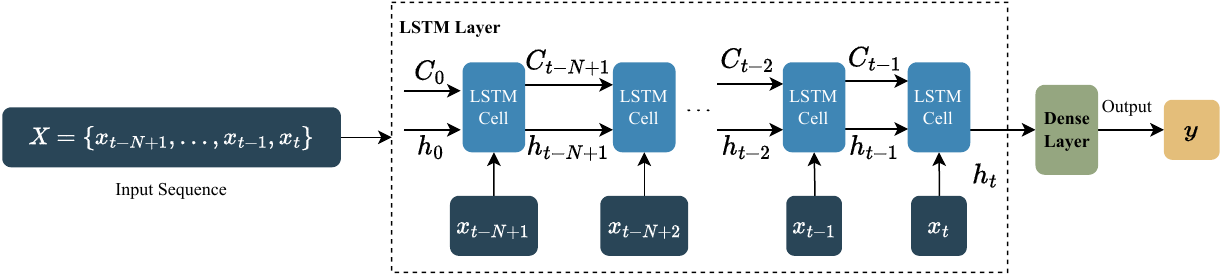}
    \caption{Unfolding the LSTM Model Architecture in the Time Dimension}
    \label{fig:Vanilla_LSTM}
\end{figure}

To better describe the iteration process, we unfold it in the time dimension (see Figure \ref{fig:Vanilla_LSTM}). Taking time step $t-1$ as an example, the LSTM cell takes the previous hidden state $h_{t-2}$ and cell state $C_{t-2}$, as well as the current input $x_{t-1}$ as input, and produces the current hidden state $h_{t-1}$ and cell state $C_{t-1}$. These states are then propagated to the next time step, allowing the model to retain contextual information. The initial values for $h$ and $C$ are typically set to 0 and denoted as $h_{0}$ and $C_{0}$. 

The LSTM cell contains three gates: input gate $i_t$, output gate $o_t$, and forget gate $f_t$, to regulate which information to keep or discard. The intermediate result $g_t$ is used to update $C_t$. Equations \ref{eq3.1} to \ref{eq3.6} represent the computations within the LSTM cell and are explained in more detail in \cite{hochreiter1997long}. The $\ast$ denotes the \textit{Hadamard} product, $[\cdot, \cdot]$ represents vector concatenation, and $W$ denotes the weight matrices of each gate \cite{qian2023enhancing}. 

\begin{align}
    i_t & =\text{Sigmoid}\left(W_{i}[h_{t-1},x_{t}]+b_{i}\right) \label{eq3.1} \\
    f_t & =\text{Sigmoid}\left(W_{f}[h_{t-1},x_{t}]+b_{f}\right) \label{eq3.2} \\
    g_t & =\text{Tanh} \left(W_{g}[h_{t-1},x_{t}]+b_{g}\right) \label{eq3.3} \\
    C_t & =f_t * C_{t-1}+i_t * g_t \label{eq3.4} \\
    h_t & =o_t * \text{Tanh} \left(C_t\right) \label{eq3.5}\\
    o_t & =\text{Sigmoid}\left(W_{o}[h_{t-1},x_{t}]+b_{o}\right) \label{eq3.6}
\end{align}

The LSTM layer outputs its newest hidden states $h_{t}$ of length $K$ when all elements in the input sequence are processed. Subsequently, the dense layer processes this output to generate the final output $y$ with $P$ dimensions. The specific task that the model needs to solve determines the exact meaning of $y$.

\section{Solution Design}
Our overall goal is to create a template-based \textit{register-transfer level} (RTL) design for FPGAs that is (a) able to support larger LSTM models with multiple cells and layers as well as large hidden size, and (b) is optimised for energy efficiency by maximising throughput. In addition, we aim to provide a flexible design that can be tailored to different usage contexts. 

To support larger LSTM models, we categorise FPGA hardware resources into critical and general, represented respectively by DSPs and LUTs. DSPs are considered critical because they are faster in executing arithmetic computations and are limited in number compared to LUTs. Our design aims to optimise the LSTM accelerator components by reducing or eliminating the use of DSPs and allocating DSPs to components that require them the most. However, this may increase the utilisation of LUTs. Therefore, we also intend to optimise the components to utilise fewer general resources, minimising the overall system resource utilisation. 

\begin{align}
   {Energy\ Efficiency}\ [GOP/s/W]  = \frac{Throughput[GOP/s] }{Power [W]} \label{eqeef}
\end{align}

In addition, energy efficiency is an essential metric for embedded applications. We employ performance per watt to measure it. As demonstrated in Equation \ref{eqeef}, \textit{throughput}, normalised by $10^9$, refers to the number of equivalent operations executed per second, and \textit{power} is the power consumption of the FPGA while running. We believe that the energy efficiency of an accelerator can be improved by increasing the throughput while consuming less power. 

\subsection{8-bit Fixed-point Quantisation}
In FPGA designs, applying fixed-point data is a common approach to balance the trade-off between precision and resource efficiency. In this work, we use the notation $(a, b)$ to represent fixed-point data, where $a$ represents the number of fractional bits (i.e. bits representing numbers smaller than 1) and $b$ represents the total width in bits. When we refer to 8-bit quantisation or 8-bit fixed-point data in the following context, we mean that $b$ is set to 8.

We observed that when using fixed-point data less than or equal to 8-bit, implementing a fixed-point multiplier with LUTs can reach a speed comparable to DSPs, which is in line with our idea of reducing or avoiding the use of DSPs. At the same time, studies have shown that 8-bit fixed-point quantisation can conserve more resources while maintaining an acceptable model precision \cite{krishnamoorthi2018quantizing,yang2020training}. Hence, although we support larger bit widths, our design uses 8-bit fixed-point quantisation as its standard. Note that although our design also supports lower bit widths, ternarisation and binarisation quantisation are difficult to use for LSTMs and have mostly been used for partially quantised LSTM models, which do not meet our needs.

\subsection{Activation Function Optimisation}
As described in Section \ref{sec:background}, the calculation of the LSTM cell necessitates the use of Tanh and Sigmoid activation functions. However, since these functions involve exponential computations, their arithmetic implementation on the FPGA can be resource inefficient and slow. One solution for this is to implement Tanh and Sigmoid with lookup tables. This avoids using DSPs and works without iterative computation. However, as demonstrated by Qian et al. \cite{qian2023enhancing}, a large lookup table with 256 entries is required to provide an acceptable precision. This can again be resource-inefficient and adds delay due to the increased logic complexity. 

As an alternative, we can replace Tanh and Sigmoid with HardTanh\footnote{\url{https://pytorch.org/docs/stable/generated/torch.nn.Hardtanh.html}} (shown in Equation \ref{eq3.8}) and HardSigmoid\footnote{\url{https://pytorch.org/docs/stable/generated/torch.nn.Hardsigmoid.html}} functions (shown in Equation \ref{eq3.7}). They have piecewise-linear characteristics, which typically take no more than two iterations for computation, requiring fewer hardware resources. Although they behave differently from Tanh and Sigmoid functions, the performance of a model using them as alternative activation functions is comparable after training \cite{manjunath2020low}. Hence, we opt to implement the HardTanh and HardSigmoid functions instead of the Tanh and Sigmoid functions. 

\begin{equation}
    \text{HardTanh(x)}=\begin{cases}
			\text{max\_val} & \text{ if  x $>$ max\_val}\\
                \text{min\_val} & \text{ if  x $<$ min\_val}\\
                \text{x}  & \text{otherwise}
		 \end{cases}
   \label{eq3.8}
\end{equation}

\begin{equation}
    \text{HardSigmoid(x)}=\begin{cases}
			\text{$0$} & \text{if x $\leq$ -3}\\
                \text{$1$} & \text{if x $\ge$ 3 }\\
                x/6 $+$ 1/2 & \text{otherwise}
		 \end{cases}
   \label{eq3.7}
\end{equation}

The HardSigmoid function in the PyTorch framework has a slope of $1/6$ for its linear interval $(-3, 3)$. Hasib-Al-Rashid et al. \cite{manjunath2020low} demonstrated that setting the slope to $1/5$ in their LSTM accelerator design yields good results. However, both $1/6$ and $1/5$ are not supported by 8-bit fixed-point data. We implement a customised HardSigmoid function with a configurable slope. To distinguish it from the HardSigmoid function in the PyTorch framework, we refer to our customised implementation as HardSigmoid$^{*}$, where its slope must be supported by the fixed-point configuration in our architecture. The slope of the HardSigmoid function is approximately 0.167. For our standard fixed-point configuration of (4,8), numbers close to 0.167 are 0.125 and 0.1875. Since 0.125 equals $1/8$, we can use bit-shifting to perform the division. Thus, in our experiments, we set the slope of the HardSigmoid$^{*}$ function to 0.125.

\subsection{ALU Optimisation}
Equations \ref{eq3.1} to \ref{eq3.6} illustrate that most computations in an LSTM cell are vector inner product calculations. Fixed-point Multiply-Accumulation (MAC) operations (see Algorithm \ref{algorithm:fixed-point-mac}) can be used to perform such calculations. Thus, optimising MAC operations with respect to speed and resource consumption is crucial.

Qian et al. \cite{qian2023enhancing} proposed an ALU that integrates lines 3-6 in Algorithm \ref{algorithm:fixed-point-mac} into a single operation. This allows the LSTM cell to perform one MAC iteration in a single clock cycle. However, this limits them to a maximum operating clock frequency of 100MHz, which cannot be increased without failing the timing requirement. Additionally, all their ALUs require the use of DSPs. One LSTM cell needs 7 of the 20 DSPs available on the \textit{XC7S15} FPGA, restricting their accelerator to support a maximum of 2 LSTM cells. 

\begin{algorithm}
    \centering
    \caption{MAC implementation for fixed-point vectors inner product}
    \label{algorithm:fixed-point-mac}
    \renewcommand{\algorithmicrequire}{\textbf{Input:}}
    \renewcommand{\algorithmicensure}{\textbf{Output:}}
    \begin{algorithmic}[1]
    \REQUIRE $W$, $x$, each of them is $N$ element vector
    \STATE Initialisation: $sum \leftarrow 0 $, $ i \leftarrow 0$
    \REPEAT
    \STATE Load $W[i]$ and $x[i]$
    \STATE $mul_{16} \leftarrow W[i]*x[i]$ \COMMENT{$mul_{16}$ is a fixed-point data in (8,16)}
    \STATE $mul_{8} \leftarrow f_{\text{round}}(mul_{16})$ \COMMENT{$mul_{8}$ is a fixed-point data in (4,8) }
    \STATE $sum \leftarrow sum + mul_{8}$
    \STATE $i \leftarrow i+1$
    \UNTIL $i=N$ 
    \ENSURE $sum$
    \end{algorithmic}
\end{algorithm}

A possible solution is to employ parallel ALUs to speed up the vector inner product calculations. For instance, if two ALUs are used for one vector inner product calculation, the time required for this calculation can be reduced by half, improving throughput. However, using more ALUs leads to additional resource consumption, further limiting the potential model size. 

A more efficient approach that does not require more ALUs is to construct a pipeline where each stage completes a single line in Algorithm \ref{algorithm:fixed-point-mac}. The stage with the highest latency determines the maximum clock frequency. 
Line 4 has the highest latency, given that multiplication is the most complex operation in the loop. Although this approach may add development overhead, it is still worth considering for embedded applications that require extreme energy efficiency.

\section{Implementation}
In this section, we present our implementation-level optimisations and design decisions. 
Firstly, we outline how we implemented the chosen activation functions and characterise the resulting performance and resource consumption. 
We then present the details of our pipelined ALU implementation, which substantially improves the maximum operating clock frequency. Finally, we present the overall resulting accelerator architecture and describe supported meta-parameters.

\subsection{Activation Function Implementation}
Based on our decision in Section 4 to replace the original activation functions, we describe the implementation details of HardTanh and HardSigmoid$^{*}$. Implementing the HardTanh function on the FPGA is straight-forward. Only two fixed-point comparators are required because the slope of its linear interval is 1 (see Equation \ref{eq3.7}). This slope value enables the implementation to maintain the same precision as the PyTorch framework, as long as the selected val\_max and val\_min are supported by our fixed-point configuration. We synthesised the HardTanh function in Vivado and found that it consumes only 5 LUTs.

The implementation of the HardSigmoid$^{*}$ function is more complex, and the best choice depends on the optimisation goal and the used quantisation.
We experimented with three methods. The first method is referred to as HardSigmoid$^{*}$-arithmetic (abbreviated as arithmetic). 
If the input is below -3 or above 3, it simply returns 0 or 1, respectively. Otherwise, the output is generated by performing a right arithmetic shift on the input and then adding a fixed-point value of 0.5. These two steps must be executed sequentially, increasing delay. The two remaining methods for implementing HardSigmoid$^{*}$ are based on lookup tables. Both produce the same behaviour as the arithmetic method. They are referred to as HardSigmoid$^{*}$-1to1 (abbreviated as 1to1) and HardSigmoid$^{*}$-step (abbreviated as step). The lookup table in the 1to1 method enumerates all input-output pairs of HardSigmoid$^{*}$. For a fixed-point configuration (4,8), this results in 96 entries.
The step method merges entries in the lookup table that have the same output. The output of HardSigmoid$^{*}$ is in $[0, 1]$. With a fixed-point configuration (4,8), only 16 output values can be represented in this range. Thus, some entries have the same output. To merge these entries, we take advantage of the monotonically increasing nature of HardSigmoid$^{*}$ and merge adjacent entries with the same output. After performing the merge operation on all entries, we obtain a step function with 14 entries. 

\begin{table}[htb]
\renewcommand\arraystretch{1.2}
\tabcolsep=0.1cm
\caption{Comparing Methods for Implementing the HardSigmoid$^{*}$ Activation Function}
\centering
\begin{tabular}{ccccc}
\specialrule{.2em}{.1em}{.1em}
\textbf{\begin{tabular}[c]{@{}l@{}}Fixed-point\\ Configuration\end{tabular}} & \textbf{Metrics} & \textbf{arithmetic} & \textbf{1to1} & \textbf{step} \\
\hline
                    & Logic Delay [ns]    & 3.765                          & 3.778                        & \textbf{ 3.660}  \\ \cline{2-5} 
\multirow{-2}{*}{(4,8)} & LUTs utilisation     & 6                              & 8                            & \textbf{ 3}     \\ \hline
                    & Logic Delay [ns]    & 5.897                          & \textbf{3.908} & 4.175                        \\ \cline{2-5}
                    
\multirow{-2}{*}{(6,8)} & LUTs utilisation     & 36                             & \textbf{27}    & 28                           \\ 
\hline
                    & Logic Delay [ns]    & 10.883                         & \textbf{4.872} & 6.360                         \\ \cline{2-5} 
\multirow{-2}{*}{(8,10)} & LUTs utilisation     & \textbf{46}      & 117                          & 1793                         \\ 
\specialrule{.2em}{.1em}{.1em}
\end{tabular}
\label{tab:hardsigmoid}
\end{table}

We compared the performance of the three methods using measures obtained from the Vivado synthesis report. The results are summarised in Table \ref{tab:hardsigmoid}. For the fixed-point configuration (4,8), we observed that the step method outperforms the others regarding resource utilisation and logic delay. This is consistent with the fact that the step method has far fewer entries than the 1to1 method. However, it is worth noting that decreasing the number of entries by 85.43\% only saves 62.5\% of LUTs because merging entries creates additional overhead for building more complex comparators.

Interestingly, for higher fractional bit widths, the situation changes. When using six fractional bits, the 1to1 method outperforms the others. The step method involves too much additional overhead. This becomes even more prominent for larger fixed-point representations. For (8,10) fixed-point configuration, the step method uses the most LUTs. The 1to1 method is the fastest. However, while being the slowest of the three methods, the arithmetic method now uses the least LUTs. As a result of these measurements, we decided to offer all three methods and let the user select one as needed. 

\subsection{Pipeline-Based ALU Implementation}

\begin{figure}[!htb]
    \centering
    \includegraphics[width=0.74\textwidth]{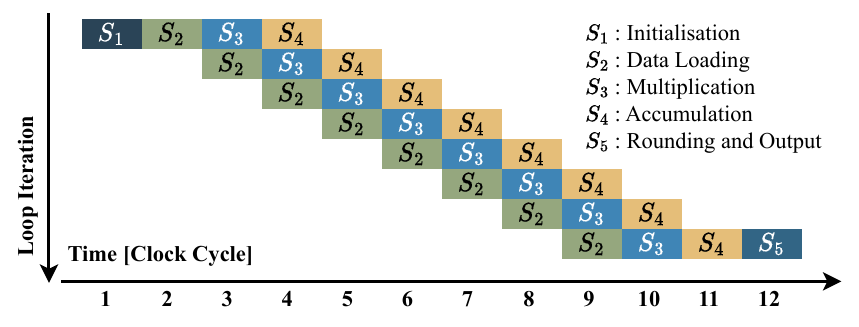}
    \caption{Pipelined Loop with Five Stages and Eight Iterations}
    \label{fig:ALU_pipeline}
\end{figure}

We constructed a pipeline-based ALU with a 5-stage depth for fixed-point MAC operation. Taking the vector length of eight as an example, as depicted in Figure \ref{fig:ALU_pipeline}, the first stage ($S_1$) involves initialisation, identical to line 1 in Algorithm \ref{algorithm:fixed-point-mac}. In the subsequent stage ($S_2$) (see line 3), two numbers from the corresponding vectors are loaded. 
In stage $S_3$ (i.e. line 4), they are multiplied, and the result is stored as 16-bit fixed-point data and propagated to the next stage. At stage $S_4$ (see line 6), the intermediate result is added to the accumulation sum. After the final iteration, in the last stage ($S_5$), the accumulation sum is rounded to 8 bits and output. Note that in contrast to Algorithm \ref{algorithm:fixed-point-mac}, this rounding is not done after each multiplication but only at the end. 

As shown in Figure \ref{fig:ALU_pipeline}, from the 4th to the 9th clock cycle, our pipeline executes 3 lines of Algorithm \ref{algorithm:fixed-point-mac} in parallel, potentially providing 3$\times$ higher throughput. Nevertheless, in the beginning (from 1st to 3rd clock cycle) and the end (from 10th to 12th clock cycle), the pipeline performs lower throughput. The longer the vector is, the higher the average throughput can be obtained with our pipeline-based ALU approach. For instance, suppose we need to calculate the dot product of 20-length vectors. In such cases, our pipeline approach can offer up to a 2.5$\times$ increase in throughput.  
However, since the multiplication stage is slower than the others, the essential throughput gain is below 2.5$\times$ in practice.

\subsection{Parameterised Architecture}
\begin{figure}[!htb]
    \centering
    \includegraphics[width=0.76\textwidth]{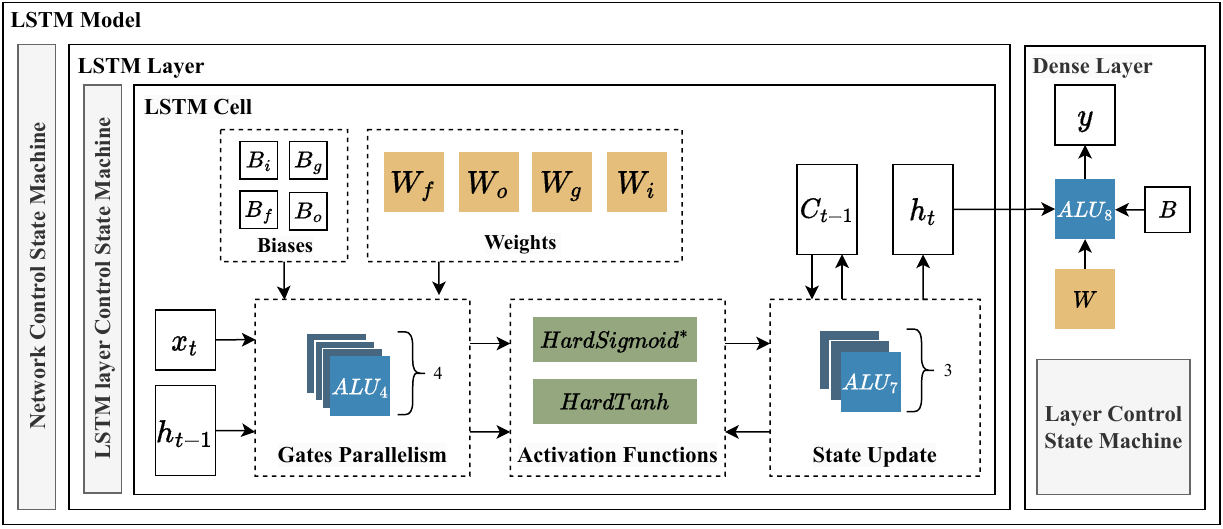}
    \caption{LSTM Accelerator Architecture Overview}
    \label{fig:arch}
\end{figure}

The overall architecture of our LSTM accelerator is shown in Figure \ref{fig:arch}. The presented LSTM model consists of (1) a single LSTM layer with a single LSTM cell and (2) a single dense layer afterwards. 
This model is also used for our experiments in Section \ref{sec:evaluation}. 
The architecture contains two parallel instances of our pipelined ALU implementation, one for $x_t$ and $h_{t-1}$, the second one for $C_{t-1}$, the two activation functions, and all weights and biases. No additional off-chip memory is needed. 
We provide a number of meta-parameters for our design (see Table \ref{table:parameters}), that can be used to adapt it to different usage contexts. Some are used to specify the functional structure of a cell or layer. As an example, \textit{hidden\_size} specifies the number of hidden units in the internal state of the LSTM cell. Others can be used to configure the implementation of the resulting accelerator. For example, \textit{ALU\_resource\_type} specifies if an ALU implementation in a LSTM cell should use DSPs or LUTs. This way, the designer can choose to save DSPs for other cells or layers in a more complex model.

Note that due to the limited number of DSPs available on the FPGA, the system prioritises allocating DSPs to ALUs on the critical path to increase the system clock frequency. This strategy is employed to make the most out of the available DSP resources. Furthermore, when selecting the \textit{weight\_resource\_type} parameter, if weights such as $W_{f}$ are assigned to BRAM-type resources, a multiple of 18 Kbit BRAM-type resources will be utilised. 

\begin{table}[htb]
\caption{Meta-Parameters of LSTM Accelerator Architecture}
\centering
\begin{adjustbox}{center}
\begin{tabular}{llcll}
\specialrule{.2em}{.1em}{.1em}
  \textbf{Meta-Parameter} & \textbf{Description}             \\ \hline
  hidden\_size (integer)  & number of the hidden units in {[}1, 200{]} \\ \hline

 input\_size (integer)  & dimension of input sample in {[}1, 10{]} \\ \hline
 
 ALU\_resource\_type (string)  & type of utilised resource of an ALU in {[}DSP, LUT{]}  \\ \hline
 
 weight\_resource\_type (string)     & \begin{tabular}[c]{@{}l@{}}
 type of utilised resource of a weights matrix\\ in {[}LUTRAM, BRAM, AUTO{]
 }\end{tabular} \\ \hline

HardSigmoid$^{*}$\_method (string)  & \begin{tabular}[c]{@{}l@{}}method of implementation of HardSigmoid$^{*}$ \\ in {[}arithmetic, 1to1, step{]}\end{tabular}    \\ \hline
 
 HardTanh\_threshold (fixed-point)   & threshold for the HardTanh implementation  \\ \hline
 
  in\_features (integer)   & size of each input sample \\ \hline
 
 out\_features (integer)   & size of each output sample  \\ 
\specialrule{.2em}{.1em}{.1em}
\end{tabular}
\end{adjustbox}
\label{table:parameters}
\end{table}

\section{Evaluation}\label{sec:evaluation}
To discuss our evaluation, we first describe the experimental settings. Then we present our evaluation results focusing on FPGA resource utilisation and throughput. Finally, we compare our power consumption and energy efficiency to related approaches.
\subsection{Experimental Settings}

To make our results comparable, we based our experiments on the study presented in \cite{qian2023enhancing}. Like them, we used the \textbf{PeMS-4W}\footnote{\url{https://doi.org/10.5281/zenodo.3939793}} dataset to predict single-step ahead traffic speed. We also adopted the LSTM model used in their study. It comprises an LSTM layer with one LSTM cell having a hidden size of 20 and a dense layer with 20 neurons. However, our design uses our replacement activation functions HardTanh (max\_val=1, min\_val=-1) and HardSigmoid$^{*}$, respectively. We also changed the quantisation, moving from (8,16) to (4,8) fixed-point configuration. We implemented and trained the modified LSTM model using the \textit{ElasticAI-Creator} \footnote{\url{https://github.com/es-ude/elastic-ai.creator}} tool. We followed the same general training settings but employed Quantisation-Aware Training instead of Post-Training Quantisation. Despite our additional optimisations, our model outperforms theirs, achieving an MSE of 0.040, which is 78\% lower than in \cite{qian2023enhancing}. 

\subsection{Resource Utilisation}
\begin{figure}[!htb]
\centering
\begin{minipage}{.485\columnwidth}
    \centering
    \includegraphics[width=1\textwidth]{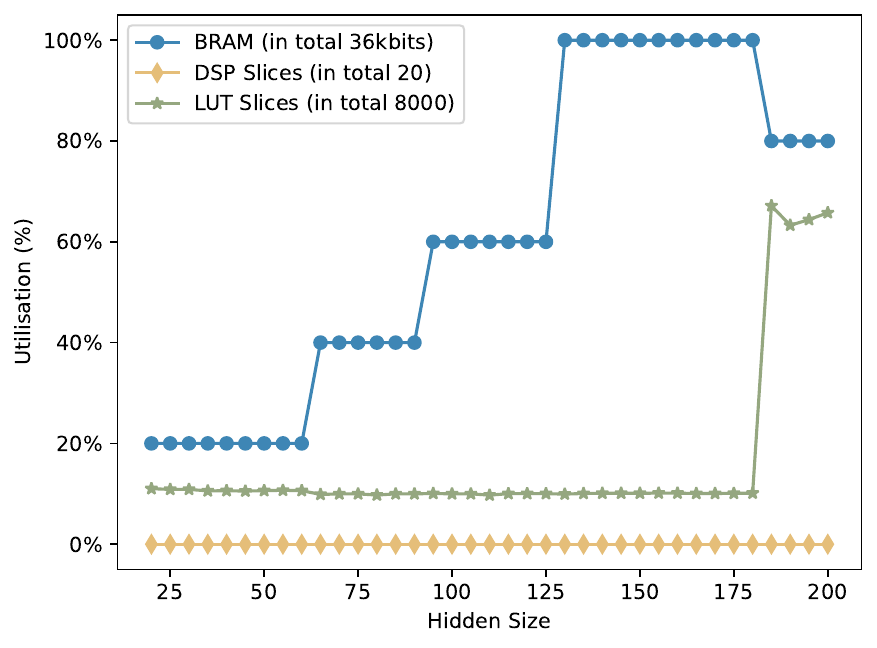}
    \caption{Utilisation without DSPs}
    \label{fig:summary_util_on_diff_hidden_size_without_dsp}
\end{minipage}
    \hspace{.01\columnwidth}
\begin{minipage}{.485\columnwidth}
    \centering
    \includegraphics[width=1\textwidth]{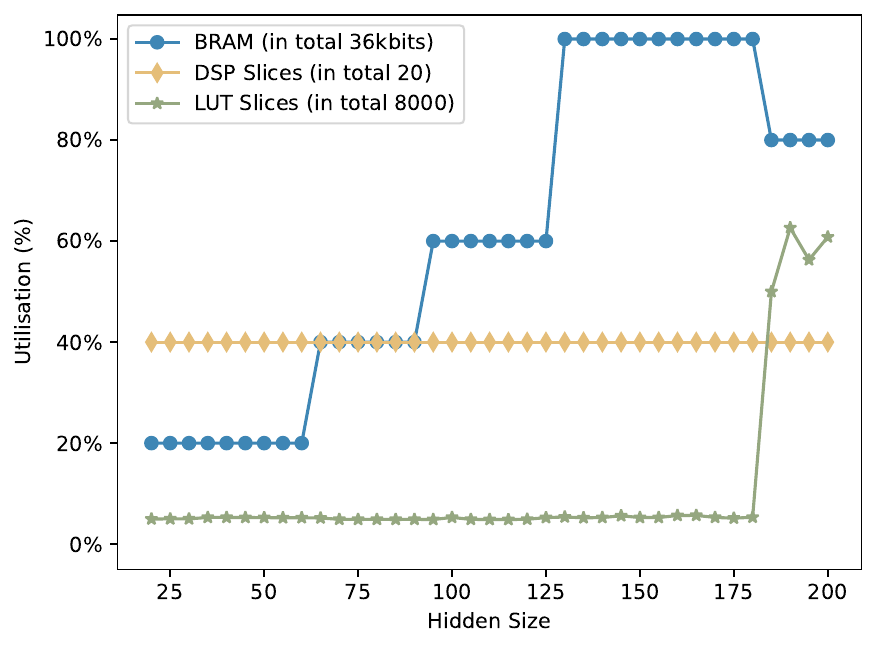}
    \caption{Utilisation with DSPs}
    \label{fig:summary_util_on_diff_hidden_size_with_dsp}
\end{minipage}%
\end{figure}

We conducted a series of experiments assessing resource utilisation to identify how complex LSTM models can be supported by our LSTM accelerator design on \textit{XC7S15} FPGA. 
Both Figures \ref{fig:summary_util_on_diff_hidden_size_without_dsp} and \ref{fig:summary_util_on_diff_hidden_size_with_dsp} show that as the hidden size of the LSTM cell increased from 20 to 200, the utilisation of BRAM, represented by the blue dotted line, changed the most significantly, which suggests that BRAM is the most critical resource to support a larger hidden size. BRAM utilisation reached a maximum of 100\% at a hidden size of 130 and remained so until the hidden size reached 180. Beyond this point, BRAM utilisation decreased, and the utilisation of LUTs increased significantly. This is because when BRAM was exhausted, Vivado switched to using LUTRAM (included in LUT Slices utilisation in Figures \ref{fig:summary_util_on_diff_hidden_size_without_dsp} and \ref{fig:summary_util_on_diff_hidden_size_with_dsp}) to implement some of the weights. Storing weights in the BRAM is preferred because it has fast access latency. Therefore, for an LSTM model with only one LSTM layer, the maximum hidden size of the LSTM cell should be 180 to ensure optimal speed on the \textit{XC7S15} FPGA.

In addition, Figures \ref{fig:summary_util_on_diff_hidden_size_without_dsp} and \ref{fig:summary_util_on_diff_hidden_size_with_dsp} were obtained under different settings of the meta-parameter \textit{ALU\_resource\_type}. In Figure \ref{fig:summary_util_on_diff_hidden_size_without_dsp}, all ALUs were set to use ``LUT" as their resource type, resulting in a constant value of 0 for the utilised DSPs. On the other hand, in Figure \ref{fig:summary_util_on_diff_hidden_size_with_dsp}, all ALUs were set to use ``DSP" as their resource type, resulting in a constant value of 40\% for the utilised DSPs, as the LSTM and dense layers occupy 8 out of the 20 available DSPs.

As we mentioned before, not utilising DSPs will inevitably increase the overhead of LUTs to realise ALUs. Comparing these two figures, we can see that the utilisation of LUTs shows a consistent difference. Before the BRAM is exhausted, the difference in LUT utilisation is between 4.375 and 6.03\%. This indicates that the LUTs consumed by implementing an 8-bit fixed-point multiplier account for at most 0.74\% of all LUTs in the \textit{XC7S15} FPGA, which is equivalent to about 60 LUTS. Based on this, we can estimate that up to five LSTM layers can be instantiated simultaneously on this FPGA when the hidden size of each cell is 60. This is especially beneficial for complex LSTM models, such as Bi-LSTM and Auto-encoders, which often require multiple LSTM layers and large hidden size. By contrast, \cite{qian2023enhancing} relied on DSPs to perform arithmetic logic in their approach, which limited their ability to implement more than two LSTM layers on this FPGA, as each layer consumes 7 of 20 available DSPs.

\subsection{Throughput}
The aim of this set of experiments is to assess the effect of different implementations of our LSTM accelerator architecture on throughput. To determine the throughput of the accelerator, we first need to obtain its maximum operating frequency by conducting timing analysis in Vivado.

\begin{table}[htb]
\centering
\caption{Frequency and Throughput for Different Optimisation Options}
\begin{adjustbox}{center}
\begin{tabular}{cccccc}
\specialrule{.2em}{.1em}{.1em}
\multirow{3}{*}{}  & \multirow{3}{*}{\textbf{\cite{qian2023enhancing}}} & \multicolumn{4}{c}{\textbf{this work$^\dagger$}} \\ \cline{3-6} 
    &     & \multicolumn{3}{c}{\textbf{\begin{tabular}[c]{@{}c@{}}HardSigmoid$^{*}$\\without Pipelined ALU\end{tabular}}}  & \multirow{2}{*}{\textbf{\begin{tabular}[c]{@{}c@{}}Pipelined ALU \\ \& step\end{tabular}}} \\ 
\cline{3-5}
    &     & \textbf{arithmetic} & \textbf{1to1} & \textbf{step} &  \\ 
\hline
\textbf{Maximal Clock[MHz]}& 100 & 104 & 109 & 115 & \textbf{204} \\ 
\hline
\textbf{Latency[$\mu$s]}        & 57.25    & 55.05  & 53.09  & 49.75  & \textbf{28.07}  \\ 
\hline
\textbf{Throughput[GOP/s]} & 0.363    & 0.378  & 0.399  & 0.417  &  \textbf{0.740}  \\ 
\hline
\textbf{Improvement}      & 1$\times$  & 1.04$\times$  & 1.09$\times$ & 1.15$\times$  &  \textbf{2.04$\times$} \\ 
\specialrule{.2em}{.1em}{.1em}
\multicolumn{6}{l}{\small $\dagger$ All implementations used the HardTanh} 
\end{tabular}
\end{adjustbox}
\label{table:perfmance_c1}
\end{table}

Table \ref{table:perfmance_c1} indicates that replacing the Tanh and Sigmoid functions with the HardTanh and HardSigmoid (Columns 2 through 4) functions resulted in a slight improvement in maximum clock frequency and accelerator performance, as compared to the work by Qian et al. \cite{qian2023enhancing} (Column 1). This is because the ALU implementation without pipeline constraint the maximum clock frequency.
The step method led to the highest increase in throughput among the three methods, at 1.15$\times$, while the arithmetic method resulted in the lowest increase, at 1.04$\times$. 
When combined with pipelined ALUs implementation (Column 5), the step method further increased the maximum clock frequency and resulted in a nearly twofold increase in throughput of $2.04\times$, along with a 50.97\% reduction in latency. It is important to note that the maximum improvement in throughput being less than 2.5$\times$ is not surprising, as multiplication takes more time than the other stages.

\subsection{Power Consumption and Energy Efficiency}

We estimated the power consumption of the accelerator at its maximum operating frequency of 204 MHz using the Xilinx Power Estimator\footnote{\url{https://www.xilinx.com/products/technology/power/xpe.html}} software. This allows us to determine the energy efficiency of the accelerator and compare it with the state-of-the-art. Table \ref{table:perfmance_c2} shows that our work achieved higher energy efficiency (Column 4) compared to the recently published work by Qian et al. \cite{qian2023enhancing} (Column 3), with a 2.33$\times$ improvement. Our proposed optimisation method achieved this, which reduced latency by 2.04$\times$ and power consumption by 1.22$\times$.

\begin{table}[htb]
\caption{Comparison with State-of-the-Art}
\centering
\tabcolsep=0.15cm
\begin{tabular}{ccccccc}
\specialrule{.2em}{.1em}{.1em}
\multicolumn{2}{c}{}                                      & \textbf{\cite{manjunath2020low}} & \textbf{\cite{chen2021eciton}} & \textbf{\cite{qian2023enhancing}} & \multicolumn{2}{c}{\textbf{this work}} \\ \hline
\multicolumn{2}{c}{\textbf{FPGA Model}}                   &  XC7A100T             &  UP5K       &     \multicolumn{3}{c}{XC7S15}                   \\ \hline
\multicolumn{2}{c}{\textbf{Utilised DSPs}}                & 4                     &      6      & 8                   & \textbf{8}                    &        \textbf{0}           \\ \hline
\multicolumn{2}{c}{\textbf{Maximal Clock{[}MHz{]}}}       & 52.6                  &     17      & 100                 & \textbf{204}                  & \textbf{204}             \\ \hline
\multirow{3}{*}{\textbf{Power{[}mW{]}}} & \textbf{Static}  & \ 92$^\dagger$      &     \ 0   &  32$^\dagger$     & 32$^\dagger$              &  32$^\dagger$          \\ \cline{2-7} 
                                       & \textbf{Dynamic} & \ 17$^\dagger$     &      17   &  38$^\dagger$     & \textbf{25}$^\dagger$      &  31$^\dagger$           \\ \cline{2-7} 
                                       & \textbf{Total}   & 109$^\dagger$     &      17   &  70$^\dagger$     & \textbf{57}$^\dagger$      &  63$^\dagger$         \\ \hline
                                       
\multicolumn{2}{c}{\textbf{Latency{[}$\mu$s{]}}}          &\multicolumn{2}{c}{incomparable}     & 53.32      & \textbf{28.07}             & \textbf{28.07}             \\ \hline
\multicolumn{2}{c}{\textbf{Energy{[}$\mu$J{]}}}           &\multicolumn{2}{c}{incomparable}     & 3.70       & \textbf{1.51}               & 1.67         \\ \hline
\multicolumn{2}{c}{\textbf{Throughput {[}GOP/s{]}}}       &    0.055 &    0.067  & 0.390       & \textbf{0.740 }              & \textbf{0.740 }           \\ \hline
\multicolumn{2}{c}{\textbf{Energy Efficiency {[}GOP/s/W{]}}}&    0.50   &    3.90    & 5.57       & \textbf{12.98}              & 11.75           \\ 
\specialrule{.2em}{.1em}{.1em}

\multicolumn{6}{l}{\small $\dagger$ Measurements come from Xilinx Power Estimator } 
\end{tabular}

\label{table:perfmance_c2}
\end{table}

Interestingly, the implementation of the ALU without DSPs (Column 5) exhibits higher dynamic power than the implementation with DSPs (Column 4), resulting in 9.47\% lower energy efficiency. Nevertheless, this approach has the advantage of not being limited by DSP resources, allowing it to support more complex LSTM models. In contrast, implementing ALUs with DSPs can be a practical choice for power efficiency applications. Moreover, we observed that using DSPs to implement all the ALUs does not lead to further increases in the maximum operating clock frequency. This is likely since using DSPs introduces net delay, as DSPs are only available in a restricted area. Consequently, the reduced logic delay achieved by using DSPs is offset by the increased net delay. 

To ensure the correctness of our values, we also measured on real hardware. The results are similar to the ones obtained by the Estimator. The average power consumption during inference when using DSPs for all ALUs is 57.4mW. Not using DSPs consumes 65.7mW. In addition, the processing time per inference for both accelerators is 2.35 $\mu$s slower than the estimated time. Our approach achieved 11.89 GOP/s/W energy efficiency on real hardware, confirming its effectiveness.

\section{Conclusion and Future Work}
Our LSTM accelerator architecture for embedded FPGAs shows that by combining 8-bit quantisation with an accompanying activation function implementation as well as with optimisations to support higher clock frequencies, we can achieve superior resource utilisation and energy efficiency compared to state-of-the-art approaches. We can reduce utilised LUTs by 29.62\% and LUTRAM by 33.33\%. Our design supports LSTM models with up to 5 layers, each with a maximum hidden size of 60, on small embedded FPGAs and allows designers to tailor accelerators to their specific needs, e.g. by choosing to get by without DSPs if needed. We can achieve a nearly 2$\times$ increase in throughput with a maximum clock frequency of 204 MHz. Power consumption is reduced by up to 18.57\% and energy efficiency per inference 59.19\%. 

In future work, we plan to verify the effectiveness of our optimised LSTM accelerator architecture in more challenging applications with bigger models. Furthermore, we plan to integrate our design into the \textit{ElasticAI-Creator} tool to enable users to generate optimised LSTM accelerators for their applications more easily. Finally, we plan to extend our work to automatically select the best parameterisation for a given context, leading to end-to-end optimisations of complex \textit{Deep Learning} models. 

\vspace{1\baselineskip}

\noindent{\textbf{Acknowledgements.}} The authors acknowledge the financial support provided by the Federal Ministry of Economic Affairs and Climate Protection of Germany in the RIWWER project (01MD22007C).

\bibliographystyle{splncs04}
\bibliography{references}

\end{document}